\documentclass[aps,prd,reprint,groupedaddress]{revtex4-2}

\usepackage{amssymb,bbold}
\usepackage{amsmath}
\usepackage{amssymb}
\usepackage[utf8]{inputenc}
\usepackage[english]{babel}
\usepackage{amssymb}
\usepackage{amsthm}
\usepackage{bbm}
\usepackage{graphicx}
\usepackage{hyperref}
\usepackage{physics}
\usepackage{systeme}
\usepackage{times}
\usepackage{epstopdf}
\usepackage{float}
\RequirePackage{color}
\RequirePackage[T1]{fontenc}
\RequirePackage{graphicx}
\RequirePackage{mathptmx}
\RequirePackage{flushend}
\newcommand{\orcid}[1]{\href{https://orcid.org/#1}
	{\includegraphics[width=7pt]{orcid.png}}}
\hypersetup{
	colorlinks=true,linkcolor=blue,citecolor=blue,
	filecolor=blue,urlcolor=blue,breaklinks=true
}

\begin{document}

\title{Exact and approximate bound state solutions of the Schr\"{o}dinger equation with a class of Kratzer-type potentials in the global monopole spacetime}

\author{Saulo S. Alves}
\email[Saulo S. Alves - ]{saulorguir@hotmail.com}
\affiliation{Departamento de F\'{\i}sica, Universidade Federal do Maranh\~{a}o, 65085-580 S\~{a}o Lu\'{\i}s, MA, Brazil}

\author{Frankbelson dos S. Azevedo}
\email[Frankbelson dos S. Azevedo - ]{frfisico@gmail.com}
\affiliation{Departamento de F\'{\i}sica, Universidade Federal do Maranh\~{a}o, 65085-580 S\~{a}o Lu\'{\i}s, MA, Brazil}

\author{Cleverson Filgueiras}
\email[Cleverson Filgueiras - ]{cleverson.filgueiras@ufla.br }
\affiliation{Departamento de F\'{i}sica, Universidade Federal de Lavras, Caixa Postal 3037, 37200-000, Lavras, Minas Gerais, Brazil}

\author{Edilberto O. Silva}
\email[Edilberto O. Silva - ]{edilberto.silva@ufma.br}
\affiliation{Departamento de F\'{\i}sica, Universidade Federal do Maranh\~{a}o, 65085-580 S\~{a}o Lu\'{\i}s, MA, Brazil}

\date{\today}
	
\begin{abstract}

This work investigates the motion of a non-relativistic charged particle within the spacetime of a global monopole. We introduce the Schrödinger equation to describe the particle's motion with two interactions by considering the Kratzer and the screened modified Kratzer potential. The problem's eigenfunctions and eigenvalues are obtained by deriving and solving the radial equation. The effective potential encompasses both the Kratzer and electrostatic self-interaction potential and leads to bound states solutions. The energy spectrum is investigated, particularly emphasizing its dependence on the system's physical parameters. The screened modified Kratzer potential and the screened self-interaction potential reveal an important role in influencing both the effective potential and the energy spectrum. Additionally, it also accommodates the existence of bound states. All these behaviors are illustrated with graphs and discussed in detail. 
\end{abstract}

\maketitle

\section{Introduction}

\label{sec:intro}

Topological defects are well-established study subjects predicted to arise in various manners. In the cosmological context, they are anticipated to form during phase transitions involving spontaneous symmetry-breaking mechanisms as the universe starts expanding and cooling down. These defects possess distinct topological charges and can be described by geometric spacetimes within the gravitational background. Such defects include global monopoles, cosmic strings, domain walls, and textures \cite{vilenkin1994cosmic}. Surprisingly, there is a close connection between these defects in Cosmology and topological defects in Condensed Matter Physics (CMP), as they share the same symmetry-breaking mechanism of formation, known as the Kibble mechanism \cite{kibble1976topology}. This connection represents a fascinating manifestation of the global signature of nature, enabling analogies between cosmological defects and those in CMP,  with the latter serving as a laboratory for gaining a deeper understanding of the physics of the universe \cite{moraes2000condensed,bowick1994cosmological}. Given our current limitations in observing the early universe due to insufficient energy capacity on Earth, these analogies are particularly fitting.

Within the realm of possibilities for the formation of topological defects, the global monopoles emerge due to the global $O(3)$ symmetry breaking that may have occurred during phase transitions in the early universe. This object has attracted significant interest among scientists in various branches of physics. Since the pioneering work of Barriola and Vilenkin \cite{PRL.1989.63.341}, who presented a weak field approximation solution describing the spacetime geometry of global monopoles, numerous studies have been conducted to explore the properties of this topological defect further. Among these works, some of them consider the presence of a charged particle in the spacetime of a global monopole and apply molecular potentials \cite{Universe.2023.9.9030132,CQG.2002.19.985,ahmed2022topological}. 

 Molecular potentials describe the interactions between atoms or molecules and serve as a fundamental framework for understanding the behavior and properties of matter at the molecular level. They are indispensable tools in the realm of physics, enabling us to delve into the complexities of molecular behavior and gain valuable insights into the fundamental forces governing matter. Examples of diverse molecular potentials and their applications include the Yukawa potential \cite{yukawa1935interaction,PRA.1971.4.1875,RP.2019.14.102409,JMM.2020.26.349,MP.2023.121.e2198617,RP.2022.39.105749}, Morse potential \cite{morse1929diatomic,CTP.2023.75.055202,IJGMMP.2350162,EPJD.202276.208,JCP.2022.157.144104,SR.2022.12.15188,fphy.2022.10.962717}, Hulthén Potential \cite{AMAFA.1942.28,APPA.2013.07.20,EPJC.2017.77.270,EPJP.2016.131.295,PLA.372.2008.4779,PLA.2019.383.3010,CTP.2013.59.679}, and Poschl-Teller potential \cite{poschl1933bemerkungen,IJP.2023.s12648-023-02676-1,PE.2023.145.115504,PRE.2022.106.024206,PRD.2022.105.084003,PE.2020.120.114029}. Another relevant molecular potential that will be the subject of our current investigation is the Kratzer potential \cite{kratzer1920ultraroten}, which describes the vibration-rotation spectra of diatomic molecules. The Kratzer potential exhibits a characteristic feature of approaching infinity as the internuclear distance approaches zero. This behavior arises from the repulsive forces between the constituent molecules within the potential. Inspired by the studies of the Kratzer potential and screened Coulomb potential and their wide-ranging applications in molecular physics, numerous new potentials known as screened modified Kratzer potentials have been proposed in the literature.  These potentials incorporate a screening exponential term that enables a direct recovery of the original Kratzer potential simply by setting the screening parameter to zero \cite{PS.2023.98.015403,IJP.2020.94.425,JLTP.2021.203.84,EPJP.2023.138.319,JLTP.2023.211.109,ikot2019eigensolution}.

In this work, we study the motion of a non-relativistic charged particle interacting via the Kratzer potential in the global monopole spacetime. We divide the work into two parts, and in both, we consider the self-interaction potential due to the topology of spacetime. First, we study the motion in the presence of the Kratzer potential in its standard form found in the literature. In the work's second part, we modify the interaction potentials through a screened deformation. The reason for studying this deformed potential is based on the study of important problems involving short-range interactions, such as those that occur between nucleons (protons and neutrons) in an atomic nucleus. The paper's organization is as follows. In Sec. \ref{sec:Pauli}, We write the Schrödinger equation with the Kratzer potential and determine the radial equation of motion. Subsequently, we solve the radial equation and find the eigenfunctions and eigenvalues of the particle. We end this section with an analysis of the physical implications due to spacetime's topology and the self-interaction potential in the particle energies. In Sec. \ref{modifiedsec}, we analyze the Schrödinger equation with a screened modified Kratzer potential and consider a screening effect on the self-interaction potential. To demonstrate the validity of the deformed potential, we make comparative sketches between the curves of the deformed Kratzer potential and the nondeformed potential for some values of the screening parameter $\delta$. By using the Frobenius method, we determine the eigenfunctions and eigenvalues. Results and discussions, including graphical illustrations of the energy spectrum as a function of the system's physical parameters, are presented in the sequence. Finally, In Sec. \ref{sec:conclusions}, we provide our conclusions. 

\section{Schr\"{o}dinger equation with Kratzer potential} 

\label{sec:Pauli}

The purpose of this section is to write the Schr\"{o}dinger equation 
to describe the motion of an electron interacting with the Kratzer
potential in the global monopole spacetime. The metric that represents this manifold is \cite{PRL.1989.63.341}
\begin{equation}
ds^{2}=-dt^{2}+\alpha ^{-2}dr^{2}+r^{2}\left( d\theta ^{2}+\sin ^{2}\theta
d\phi ^{2}\right) ,  \label{metric}
\end{equation}
where $\alpha ^{2}=1-8\pi G\mu ^{2}$ is smaller than $1$ and represents the deficit solid of this manifold. The parameter $\mu$ corresponds to the scale of gauge-symmetry breaking \cite{PhysRevD.95.104012}. For this spacetime, it is known that the area of a sphere of unit radius is not $4\pi r^{2}$ but $4\pi \alpha^{2}r^{2}$. 
Furthermore, another known characteristic is that the surface with $\theta=\pi/2$ presents the geometry of a gauge cosmic string with a deficit angle $\Tilde{\Delta}= 8\pi^{2}\mu^{2}$.  
It is known that the motion of massive or charged particles in the spacetime (\ref{metric}) involves the effect of the self-interaction potential in the model description. In this case, the relevant equation is the Schr\"{o}dinger equation in spherical polar coordinates with vector
coupling \cite{EPJC.2015.75.321,EPJC.2019.79.596}. The Schr\"{o}dinger equation has the form
\begin{equation}
-\frac{\hbar ^{2}}{2M}\nabla ^{2}\Psi \left( r,\theta ,\phi \right) +V\left(
r\right) \Psi \left( r,\theta ,\phi \right) =E\Psi \left( r,\theta ,\phi
\right) ,  \label{S1}
\end{equation}
where $M$ is the mass of the particle, and
\begin{equation}
V\left( r\right) =V_{SI}\left( r\right) + V_{K}\left( r\right)  \label{vr}
\end{equation}
is the effective potential, which contains the Kratzer potential $V_{K}\left( r\right)$ and the electrostatic self-interaction potential $V_{SI}\left( r\right)$.
The self-interaction potential $V_{SI}\left( r\right) $ is of the Coulomb type, and has the
following representation \cite{PRD.1997.56.1345}:
\begin{equation}
V_{SI}(r)=\frac{\mathcal{K}\left( \alpha \right) }{r},  \label{si}
\end{equation}%
where $r$ is the distance from the electron to the monopole and $\mathcal{K}%
(\alpha )$ is the constant of coupling. For this potential, depending on the
sign of $\mathcal{K}(\alpha )$, it can be attractive or repulsive. Since we
are interested in solving the model for electrostatic
interactions, the constant $\mathcal{K}(\alpha )$ is given by \cite%
{PRD.1997.56.1345} 
\begin{equation}
	\mathcal{K}\left( \alpha \right) =\frac{e^{2}S(\alpha )}{2}>0,  \label{ic}
\end{equation}%
where $e$ is the electron charge. The function $S(\alpha )$ in Eq. (\ref{ic}%
) is given by 
\begin{equation}
S(\alpha )=\sum\limits_{l=0}^{\infty }\left[ \frac{2l+1}{\sqrt{4l\left(l+1\right) +\alpha ^{2}}}-1\right] ,  \label{sb}
\end{equation}%
where $l\in\mathbb{Z}$ denotes the angular-momentum quantum number. The function $S(\alpha )$ is
a finite positive number for $\alpha <1$ and negative for $\alpha >1$. For
our purposes, to ensure the validity of equation (\ref{ic}), we only consider the case with $\alpha <1$. The Kratzer potential $V_{K}\left(r\right) $ is given by \cite{kratzer1920ultraroten,flugge1974practical}%
\begin{equation}
	V_{K}\left( r\right) =-2D\left( \frac{A}{r}-\frac{A^{2}}{2r^{2}}\right) ,
	\label{ek}
\end{equation}%
where $A$ and $D$ are positive constants.
To solve Schr\"{o}dinger equation (\ref{S1}), we proceed in the usual way: we
begin by looking for solutions that are separable into products in the form $\Psi \left( r,\theta,\phi \right) =R\left( r\right) Y\left( \theta,\phi \right)$, where the angular wave functions $Y\left( \theta,\phi \right)$
are called spherical harmonics. In addition, we must search for solutions energy eigenstates $\Psi \left( r,\theta
,\phi \right)$ where
\begin{align}
\mathbf{L}^{2}Y\left( \theta ,\phi \right) &=\hbar ^{2}l\left( l+1\right)
Y\left( \theta ,\phi \right) ,\text{with }l\in \mathbb{Z}, \,\text{and}\\
L_{z}Y\left( \theta ,\phi \right) &=\hbar mY\left( \theta ,\phi \right) ,
\text{with}\;m=-l,-l+1\ldots ,l-1,l,
\end{align}
with $\mathbf{L}$ being the usual orbital angular momentum operator in
spherical polar coordinates. The effects of applying the angular momentum
operator on the eigenfunctions become more evident through substitution $%
R\left( r\right) =r^{-1}\psi \left( r\right) $, which leads us to the radial
equation 
\begin{equation}
-\frac{\alpha ^{2}\hbar ^{2}}{2M}\psi ^{\prime \prime }\left( r\right) +%
\frac{\hbar ^{2}}{2M}\frac{l\left( l+1\right) }{r^{2}}\psi \left( r\right)
+V\left( r\right) \psi \left( r\right) =E\psi \left( r\right) ,  \label{er}
\end{equation}%
 With the inclusion of the
potentials $V_{SI}\left( r\right)$ and $V_{K}\left( r\right)$ in Eq. (\ref{er}), the radial equation takes the form%
\begin{equation}
-\frac{\alpha ^{2}\hbar ^{2}}{2M}\psi ^{\prime \prime }\left( r\right)
+V_{eff}(r)\psi \left( r\right) =E\psi \left( r\right) ,  \label{r2}
\end{equation}%
where%
\begin{equation}
V_{eff}(r)=\frac{\hbar ^{2}}{2M}\left( \frac{l\left( l+1\right) +DA^{2}}{r^{2}}\right) +\frac{\mathcal{K}\left( \alpha \right) -2DA}{r}
\label{veffND}
\end{equation}%
is the effective potential. Defining the new variable $\rho =2K_{b}r$, Eq. (\ref{r2}) can be written as%
\begin{equation}
	\psi ^{\prime \prime }\left( \rho \right) -\frac{j^{2}}{\rho ^{2}}\psi
	\left( \rho \right) -\frac{\zeta ^{\prime }}{\rho }\psi \left( \rho
	\right) -\frac{1}{4}\psi \left( \rho \right) =0,  \label{edok1}
\end{equation}%
where 
\begin{align}
	j& =\sqrt{\frac{l\left( l+1\right) }{\alpha ^{2}}-\frac{2MDA^{2}}{\alpha
		^{2}\hbar ^{2}}}, \label{j} \\ 
	\zeta ^{\prime }& =\frac{\zeta }{K_{b}},\;\;\zeta =\frac{M}{\alpha
		^{2}\hbar ^{2}}\left[ \mathcal{K}\left( \alpha \right) -2AD\right] , \label{zeta} \\
	K_{b}& =\sqrt{-\frac{2ME}{\alpha ^{2}\hbar ^{2}}}>0.  \label{K2}
\end{align}%
The Subscript $b$ in Eq. (\ref{K2}) stands for bound states. Now we must examine the asymptotic behavior of the differential equation (%
\ref{edok1}). When $\rho \rightarrow \infty $, the constant term in brackets
is dominant, and the resulting equation is%
\begin{equation}
	\psi ^{\prime \prime }\left( \rho \right) -\frac{1}{4}\psi \left( \rho
	\right) =0,  \label{edoak1}
\end{equation}%
whose general solution is given by%
\begin{equation}
	\psi \left( \rho \right) =A_{1}e^{-\frac{\rho }{2}}+A_{2}e^{\frac{\rho }{2}}.
	\label{sol1}
\end{equation}
Analyzing (\ref{sol1}), we find that $e^{\rho }$ diverges when $\rho
\rightarrow \infty $, so $A_{2}=0$. Therefore, the relevant solution results%
\begin{equation}
	\psi \left( \rho \right) \simeq A_{1}e^{-\frac{\rho }{2}}. \label{sexp}
\end{equation}%
On the other hand, when we investigate the solution (\ref{edok1}) for $\rho
\rightarrow 0$, the centrifugal term dominates, and the resulting equation is%
\begin{equation}
	\psi ^{\prime \prime }\left( \rho \right) -\frac{j^{2}}{\rho ^{2}}\psi
	\left( \rho \right) =0,  \label{sol2}
\end{equation}%
with the general solution given by%
\begin{equation}
	\psi \left( \rho \right) =B_{1}\,\rho ^\ell+B_{2}\,\rho^{1-\ell},
\end{equation}
where $\ell =(1+\sqrt{4j^{2}+1})/2$. Note that as $4j^{2}+1>1$, so $\rho^{1-\ell}$ diverges when $\rho \rightarrow 0$. From Eq. (\ref{j}), we can note that $l=0$ does not satisfy the constraint for $j$, since $l(l+1)>2MDA^2/\hbar^2$ with $D$ and $A$ being positive constants. We can take $B_{2}=0$ in this case. Therefore, the resulting solution is 
\begin{equation}
\psi \left(\rho \right) \simeq B_{1}\,\rho^{\ell}. \label{as2}
\end{equation}
In possession of the solutions (\ref{sexp}) and (\ref{as2}), we can introduce the new function 
\begin{equation}
	\psi \left( \rho \right) = e^{-\frac{\rho }{2}}\rho ^{\ell }F\left(
	\rho \right),  \label{a1}
\end{equation}
where $F(\rho)$ is a function to be determined. Substituting Eq. (\ref{a1}) into Eq. (\ref{edok1}), we obtain the differential equation
\begin{equation}
\rho F^{\prime \prime }+\left(2\ell -\rho \right) F^{\prime}-\left( \ell
+\zeta^{\prime}\right) F=0. \label{radial}
\end{equation}%
Equation (\ref{radial}) is a confluent hypergeometric differential equation,
and its solution is given by%
\begin{equation}
	F\left( \rho \right) =C_{\ell }~_{1}F_{1}\left( \ell +\zeta ^{\prime
	},2\ell ,\rho \right) ,
\end{equation}%
and then the solution (\ref{a1}) is written as%
\begin{equation}
	\psi _{n\ell }\left( \rho \right) =C_{n\ell }\,e^{-\frac{\rho }{2}}\rho
	^{\ell }~_{1}F_{1}\left( -n,2\ell ,\rho \right) ,
\end{equation}%
where $n=\ell +\zeta ^{\prime }$ and $C_{n\ell }$ is the normalization
constant. The energy eigenvalues are determined from the condition $\ell
+\zeta ^{\prime }=-n$, and making use of the Eqs. (\ref{zeta}) and (\ref{K2}). We find the expression
\begin{equation}
E_{nl}=-\frac{\alpha ^{2}\hbar ^{2}}{2M}\frac{\zeta ^{2}}{\left( n+\ell \right)^{2}},\label{selfv}
\end{equation}
where $\ell$ is given in Eq. (\ref{as2}).
\begin{figure}[!t]
\centering
\includegraphics[width=\columnwidth]{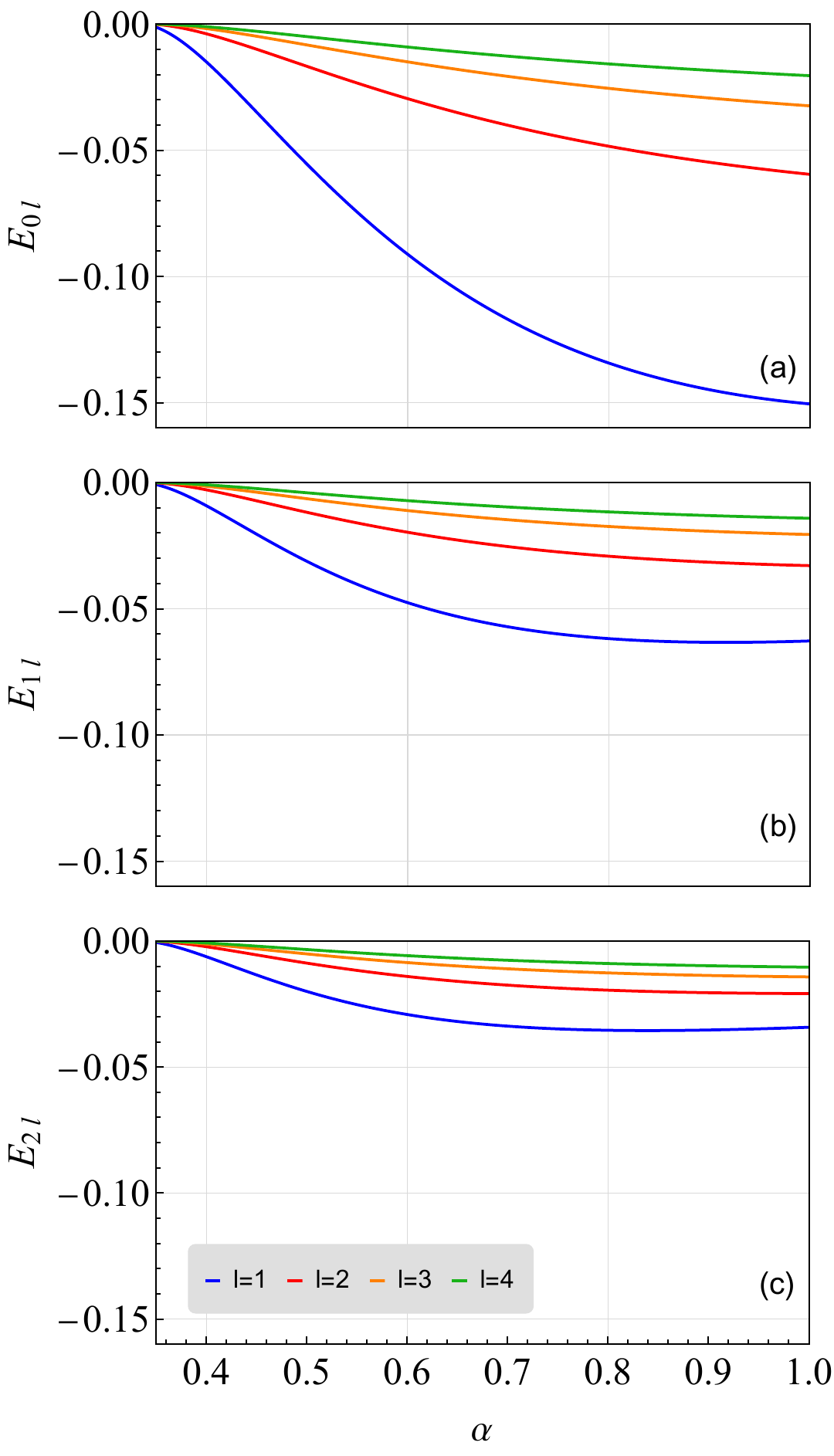}
\caption{Energy levels as a function of the monopole parameter $\alpha$ for different values of $l$. In (a), for $n=0$, (b) for $n=1$, and (c) $n=2$.
In both cases, we using $A=0.5$, $D=1$  $q=1$, $\hbar=1$ and $M=1$}
\label{f1}	
\end{figure}

In Figure \ref{f1}, it can be observed that the energy undergoes significant changes concerning the monopole parameter $\alpha$, progressively becoming more negative. $E_{01}$ represents the lowest energy level, while higher levels display fewer negative values. As the quantum number $n$ increases, the energy levels become more compact. Additionally, there is a tendency for the energy to stabilize as $\alpha$ increases. However, this stabilization effect is not yet prominent in the lowest energy levels (with $n=0$).

\section{Schr\"{o}dinger equation with screened modified Kratzer potential}
\label{modifiedsec}

In this section, we address the problem discussed in the previous section by introducing a screening deformation to the potentials (\ref{si}) and (\ref{ek}). Firstly, we intuitively assume that the self-interaction potential (\ref{si}) follows the same deformation rule as the Coulomb potential. Consequently, our proposed potential is given by:
\begin{eqnarray}
\mathcal{V}_{SI}\left( r\right) &=&\frac{\mathcal{K}\left( \alpha \right)
e^{-\delta r}}{r}, \\
\mathcal{V}_{K}\left( r\right) &=&-2D\left( \frac{A}{r}e^{-\delta r}-\frac{%
A^{2}}{2r^{2}}e^{-2\delta r}\right) \label{modified},
\end{eqnarray}%
where $\delta $ is the screening parameter. As $\delta $ approaches zero, we recover the self-interaction (\ref{si}) and Kratzer potential (\ref{ek}), respectively. The potential (\ref{modified}) differs slightly from the screened modified Kratzer potential proposed in Ref. \cite{ikot2019eigensolution}. Here, we chose to add the exponent $2$ in the second term and second exponential for simplicity in solving the equation of motion and as a way to get the inverse quadratic Yukawa potential \cite{hamzavi2012approximate}. It is important to note that the potential (\ref{modified}) is a generalized potential that incorporates the Kratzer, Yukawa (screened Coulomb), and Coulomb potentials.

With these potentials at hand, the differential equation to be solved is
\begin{equation}
-\frac{\alpha ^{2}\hbar ^{2}}{2M}\psi ^{\prime \prime }\left( r\right) +%
\mathcal{V}_{eff}\left( r\right) \psi \left( r\right) =\mathcal{E}\psi \left( r\right),
\label{Eef}
\end{equation}%
where%
\begin{align}
\mathcal{V}_{eff}\left( r\right) =\;&\frac{\hbar ^{2}}{2M}\frac{l\left(
l+1\right) }{r^{2}}
-2D\left( \frac{A}{r}e^{-\delta r}-\frac{A^{2}}{2r^{2}}e^{-2\delta
r}\right)\notag \\
&+\frac{\mathcal{K}\left( \alpha \right) e^{-\delta r}}{r}  \label{pt}
\end{align}%
is the screened effective potential. 
\begin{figure}[!t]
\centering
\includegraphics[width=\columnwidth]{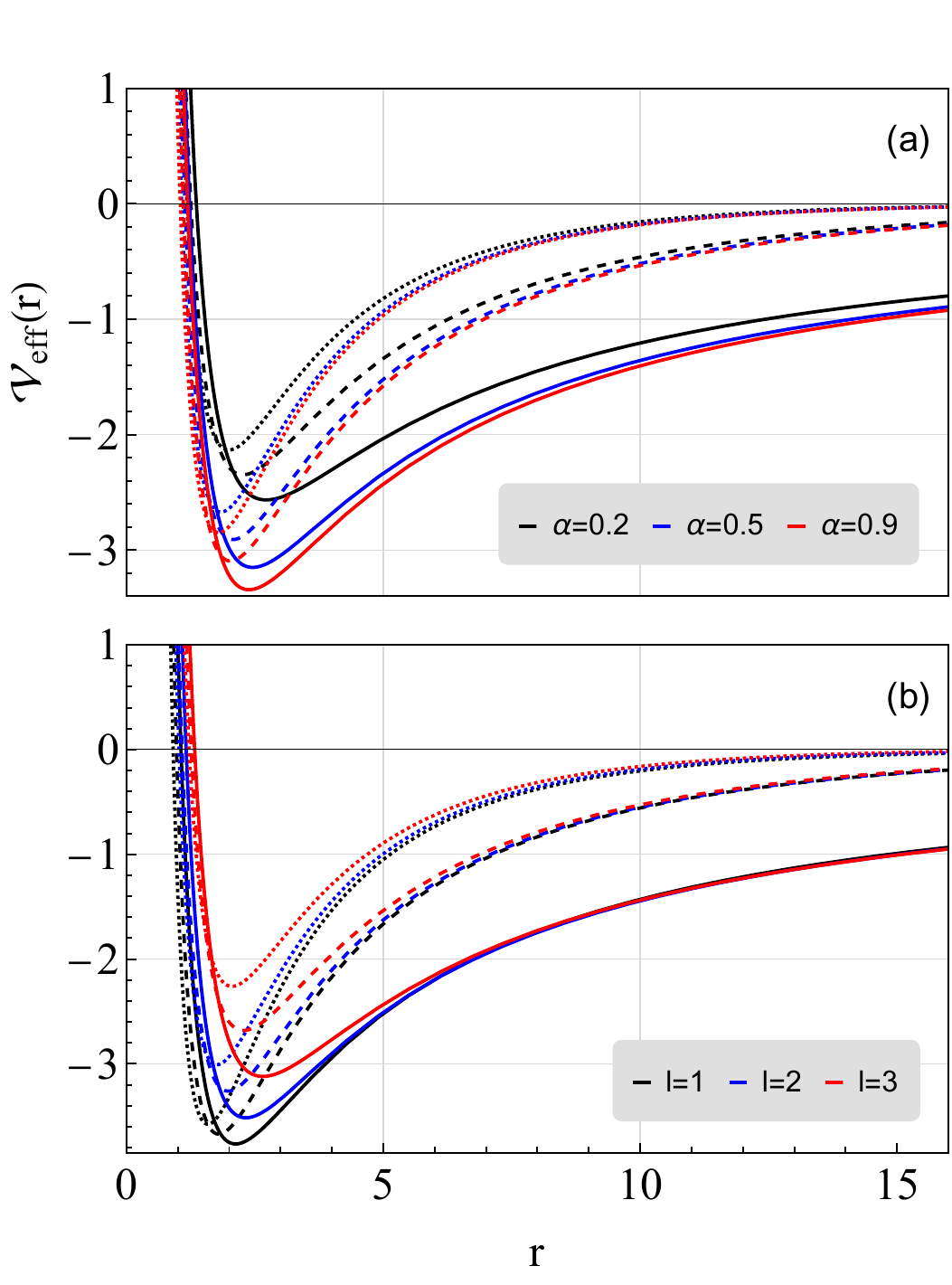}
\caption{Effective potential (\ref{pt}) as a function of $r$. The solid line is for $\delta=0$ (which leads to the effective potential (\ref{veffND})), the dashed line is for $\delta=0.1$, and the dotted line is for $\delta=0.2$. In (a), we set $l=2$ and use different values of $\alpha$. In (b), we set $\alpha=0.5$ and use different values of $l$. In both cases, we are using $A=2$ and $D=4$.}
\label{Potential}	
\end{figure}
Figure \ref{Potential} shows the effective potential given in equation (\ref{pt}) and compares it with the case where $\delta=0$, which leads to equation (\ref{veffND}). Curves with the parameter $\delta>0$ refers to the modified Kratzer potential. From Figure \ref{Potential}(a), we see that the potential well becomes deeper when the parameter $\alpha$ value increases. On the other hand, in Figure \ref{Potential}(b), we see that the effect is inverse with the increasing of the quantum angular momentum $l$. In all cases, the solid line ($\delta=0$) has a deeper well, and the increase of $\delta$ shifts the minimum potential towards the origin. 

The differential equation (\ref{Eef}) can be written in a more compact form
as%
\begin{align}
-&\psi ^{\prime \prime }\left( r\right) +\frac{\lambda ^{2}}{r^{2}}\psi
\left( r\right) +\lambda _{K}^{2}\frac{e^{-2\delta r}}{r^{2}}\psi \left(r\right) +2 \zeta\, \frac{e^{-\delta r}}{r}\psi
\left( r\right) \notag \\ 
&+k_{b}^{2}\psi \left( r\right) =0,\label{diffd}
\end{align}%
where
\begin{align}
\lambda ^{2} &=\frac{l\left( l+1\right) }{\alpha ^{2}},\; \lambda_{K}^{2} &=\frac{2 M D A^{2}}{\alpha ^{2}\hbar ^{2}},\;k_{b}=%
\sqrt{-\frac{2M\mathcal{E}}{\alpha ^{2}\hbar^{2}}}>0, \label{seq2}
\end{align}%
and $\zeta$ is given in Eq. (\ref{zeta}).
To solve the Eq. (\ref{diffd}) for $\lambda ^{2}>0$ and $\lambda _{K}^{2}>0$, we use the approximate forms 
\begin{equation}
\frac{1}{r^{2}}\simeq 4\delta ^{2}\frac{e^{-2\delta r}}{\left(
1-e^{^{-2\delta r}}\right) ^{2}},\;\; \frac{1}{r}\simeq 2\delta 
\frac{e^{-\delta r}}{1-e^{-2\delta r}},
\end{equation}
which leads to the differential equation
\begin{align}
-\psi ^{\prime \prime }\left( r\right)&+4\delta \zeta\, \frac{e^{-2\delta r}}{r}\psi \left( r\right) \notag \\&+4\delta ^{2}\lambda _{K}^{2}\,\frac{e^{-4\delta r}}{\left( 1-e^{^{-2\delta r}}\right) ^{2}}\psi \left( r\right)+k_{b}^{2}\psi
\left( r\right) =0.  \label{edo}
\end{align}%
Defining the new the variable $y=1-e^{-2\delta r}$, Eq. (\ref{edo}) takes the form%
\begin{align}
\psi^{\prime \prime }\left( y\right) &-\frac{1}{\left( 1-y\right) }\psi
^{\prime }\left( y\right) -\frac{\lambda ^{2}}{y^{2}\left( 1-y\right) }\psi
\left( y\right) -\frac{\lambda _{K}^{2}}{y^{2}}\psi \left( y\right) \notag \\
 &-\frac{\eta ^{2}}{y\left( 1-y\right) }\psi \left( y\right) -\frac{k^{2}}{4\delta^{2}\left( 1-y\right) ^{2}}\psi \left( y\right) =0 ,\label{edo2}
\end{align}%
where have defined $\eta ^{2}=\zeta
/\delta $. It can be shown that Eq. (\ref{edo2}) has regular singularities
at points $0$, $1$ and $\infty $. As will be shown following, Eq. (\ref{edo2}) is a hypergeometric-type equation. We must solve this equation using the boundary conditions%
\begin{align}
\psi \left( y=0\right) & =0\;\;\;(r\rightarrow \infty ), \\
\psi \left( y=1\right) & =0\;\;\;(r=0).
\end{align}%
Thus, we can use the Frobenius method to solve it. This can be accomplished
using a solution of the form%
\begin{equation}
\psi \left( y\right) =y^{\gamma }\left( 1-y\right) ^{\nu }\mathcal{F}\left(	y\right), \label{t1}
\end{equation}
where $\nu $ and $\gamma $ are arbitrary constants to be determined.  and
where $\mathcal{F}\left(y\right)$ is a function to be found. Note that the solution (\ref{t1}) is
finite at regular singular points $y=0$, $y=1$, and $y=\infty $.
Substituting the solution (\ref{t1}) into Eq. (\ref{edo2}) leads to the
differential equation%
\begin{align}
\mathcal{F}^{\prime \prime }\left( y\right) &+\left[ \frac{2\gamma -\left(
2\gamma +2\nu +1\right) y}{y\left( 1-y\right) }\right] \mathcal{F}^{\prime}\left( y\right) +\frac{\nu ^{2}-\frac{k_{b}^{2}}{4\delta ^{2}}}{\left(
1-y\right) ^{2}}\mathcal{F}\left( y\right)   \notag \\
-&\left[ \frac{\frac{1}{2\delta }\left( \beta _{M}-\beta _{K}\right) +2\gamma \nu +\gamma ^{2}+\lambda _{K}^{2}}{y\left( 1-y\right) }\right] \mathcal{F} \left( y\right) \notag\\&+\frac{\gamma \left( \gamma -1\right) -\lambda_{K}^{2}-\lambda ^{2}}{y^{2}\left( 1-y\right) }\mathcal{F}\left( y\right)=0. \label{ehg}
\end{align}
The differential equation (\ref{ehg}) can be written as the canonical hypergeometric differential equation. For this to be accomplished, we need to determine the values of $\nu$ for which the numerator of the third term is zero and, subsequently, find the values of $\gamma$ for which the fifth term is zero. For the first case, we have 
\begin{equation} 
\gamma \left( \gamma -1\right) -\lambda _{K}^{2}-\lambda ^{2}=0,
\end{equation}
which provides the values
\begin{eqnarray}
\gamma _{1} &=&\frac{1}{2}+\frac{1}{2}\sqrt{4\lambda ^{2}+4\lambda _{K}^{2}+1}=d, \label{gamma1}\\
\gamma _{2} &=&\frac{1}{2}-\frac{1}{2}\sqrt{4\lambda ^{2}+4\lambda _{K}^{2}+1}=1-d.\label{gamma2}
\end{eqnarray}%
For the second case, we have
\begin{equation}
	\nu ^{2}-\frac{k_{b}^{2}}{4\delta ^{2}}=0,
\end{equation}%
and the values of $\nu$ are given by
\begin{eqnarray}
\nu _{1} &=&+\frac{k_{b}}{2\delta }=+k_{b}^{\prime }, \label{nu1}\\
\nu _{2} &=&-\frac{k_{b}}{2\delta }=-k_{b}^{\prime }.\label{nu2}
\end{eqnarray}
By using $\gamma_{1}$ and $\nu_{1}$, Eq. (\ref{ehg}) takes the form
\begin{align}
y\left( 1-y\right) \mathcal{F}\left(y\right)^{\prime \prime}&+ \left[2\gamma_{1}-\left( 2\gamma _{1}+2\nu_{1}+1\right) y \right] \mathcal{F}\left(y\right)^{\prime} \notag \\& -\left[ \eta^{2}+2\gamma _{1}\nu _{1}+\gamma _{1}^{2}+\lambda _{K}^{2}\right] \mathcal{F}\left(y\right)=0. \label{ehg2}
\end{align}
Equation (\ref{ehg2}) can be solved via the Frobenius method. Before doing this, it is convenient to set the parameters
\begin{align} 
\eta _{1} &=\gamma_{1}+\nu_{1}-\sqrt{\nu _{1}^{2}+\lambda_{K}^{2}-\eta^{2}}, \\
\eta_{2} &=\gamma_{1}+\nu _{1}+\sqrt{\nu_{1}^{2}+\lambda _{K}^{2}-\eta^{2}},
\end{align}  
from which we can find 
\begin{align}
\eta _{1}+\eta_{2}&=2\gamma _{1}+2\nu _{1}, \\
\eta _{1}\eta_{2} &=-\left( \lambda _{K}^{2}-\eta ^{2}-\gamma _{1}^{2}-2\nu _{1}\gamma_{1}\right).
\end{align}%
For convenience, we also define the parameter
\begin{equation}
\eta _{3}=2\gamma _{1}.
\end{equation}%
Therefore, once the parameters $\eta_1$, $\eta_2$, and $\eta_3$ are identified, we can see that Eq. (\ref{ehg2}) is a hypergeometric differential equation of the form
\begin{align}
y\left( 1-y\right)&\mathcal{F}\left(	y\right)^{\prime \prime } +\left[ \eta
_{3}-\left( 1+\eta _{1}+\eta _{2}\right) y\right] \mathcal{F}\left(	y\right)^{\prime }\notag \\ &-\eta _{1}\eta _{2}\mathcal{F}\left(	y\right) =0.  \label{ehp}
\end{align}
Equation (\ref{ehp}) has the same form as the canonical hypergeometric equation, and its solution is well known. Let's then consider the series solution of the form
\begin{equation}
\mathcal{F}\left(	y\right) =\sum\limits_{p=0}^{\infty}a_{p}y^{p+c},~\text{with}\; a_{0}\neq 0. \label{sh}
\end{equation}
Substituting this solution into Eq. (\ref{ehp}), we get
\begin{align}
&a_{0} \left[ c\left( c-1\right) +\eta_{3} c\right] y^{c-1}+\sum
\limits_{p=1}^{\infty }a_{p}\left( p+c\right) \left( p+c-1\right)
y^{p+c-1}\notag \\
&-\sum\limits_{p=1}^{\infty }a_{p-1}\left( p+c-1\right) \left(p+c-2\right) y^{p+c-1} \notag \\
&-\left(1+\eta_{1} +\eta_{2} \right) \sum\limits_{p=1}^{\infty}a_{p-1}\left(
p+c-1\right) y^{p+c-1}\notag \\
&+\eta_{3} \sum\limits_{p=1}^{\infty}a_{p}\left( p+c\right) y^{p+c-1}-\eta_{1} \eta_{2} \sum\limits_{p=1}^{\infty}a_{p-1}y^{p+c-1}=0. \label{sre}
\end{align}
All coefficients must be zero from the linear independence of all powers of $y$. Therefore, from the first term in Eq. (\ref{sre}),
we have
\begin{equation}
a_{0}\left[ c\left( c-1\right) +\eta_{3} c\right] =0,
\end{equation}%
which solved for $c$ provides
\begin{eqnarray}
c_{1} &=&0, \label{sc1}\\
c_{2} &=&1-\eta_{3}. \label{sc2}
\end{eqnarray}%
The remaining terms in Eq. (\ref{sre}) give us the recurrence relation
\begin{equation}
a_{p+1}=\frac{\left( p+c\right) \left( p+c+\eta_{1} +\eta_{2} \right) +\eta_{1}
\eta_{2} }{\left(p+c+1\right) \left( p+c+\eta_{3} \right) }a_{p},\text{ for}\;	p\geqslant 0.  \label{rr}
\end{equation}
For the values of $c$ in Eqs. (\ref{sc1}) and (\ref{sc2}), we find the general solution
\begin{align}
\mathcal{F}\left(y\right)& = A\;_{2}F_{1}\left( \eta_{1} ,\eta_{2} ,\eta_{3} ;y\right)\notag \\
&+B\;y^{1-\eta_{3}}\;_{2}F_{1}\left( \eta_{1} +1-\eta_{3},\eta_{2} +1-\eta_{3}
,2-\eta_{3}; y\right).  \label{sgl}
\end{align}
Using the results (\ref{gamma1}), (\ref{nu1}) and (\ref{sgl}), the solution (\ref{t1}) can be written as
\begin{align}
&\psi \left( y\right)=A\,y^{\eta_{3}}\left( 1-y\right) ^{\frac{k_{b}}{2\delta }
}{}_{2}F_{1}\left( \eta _{1},\eta _{2},2d;y\right) +B\,y^{\eta_{3}}\left(
1-y\right) ^{\frac{k_{b}}{2\delta }}\notag \\& \times y^{1-2d}\, _{2}F_{1}\left( \eta_{1}+1-2d,\eta_{2}+1-2d,2-2d;y\right). \label{solA}
\end{align}
By certifying that $\eta_{3}=2d$, with $d= (1+ \sqrt{4\lambda^{2}+4\lambda_{K}^{2}+1})/2$, the solution (\ref{solA}) takes the form
\begin{align}
\psi \left( y\right)& =A\,y^{d}\left( 1-y\right) ^{\frac{k_{b}}{2\delta }}\, _{2}F_{1}\left( \eta _{1},\eta _{2},2d;y\right) +B\,y^{1-d}\left(1-y\right) ^{\frac{k_{b}}{2\delta }} \notag \\ & \times\, _{2}F_{1}\left( \eta _{1}+1-2d,\eta_{2}+1-2d,2-2d;y\right), \label{ns}
\end{align}
where
\begin{align}
\eta _{1} &=d+k_{b}^{\prime}-\sqrt{k_{b}^{\prime 2}+\lambda_{K}^{2}-\eta^{2}},\\
\eta _{2} &=d+k_{b}^{\prime}+\sqrt{k_{b}^{\prime 2}+\lambda_{K}^{2}-\eta^{2}}.
\end{align}
Since $c_2=1-\gamma \neq \mathbb{Z}$ (see Eq. (\ref{sc2})), which is a necessary condition for the convergence of the wave function, we take $B=0$ in Eq. (\ref{ns}). Hence, the physically acceptable solution is
\begin{equation}
\psi \left(y\right) = A~y^{d}\left(1-y\right) ^{\frac{k_{b}}{2\delta}}\, _{2}F_{1}\left( \eta_{1},\eta_{2},2d;y\right), \label{ff}
\end{equation}
which written to the variable $r$ reads
\begin{equation}
\psi\left( r\right) =C_{nm}~\left(1-e^{-2\delta r}\right) ^{d}e^{-\kappa
_{b}r}{}_{2}F_{1}\left( \eta _{1},\eta _{2},2d;1-e^{-2\delta r}\right).\label{nsol}
\end{equation}
If $\eta_{1}=-n$, with $n\in \mathbb{Z}$, the series $_{2}F_{1}\left( \eta_{1},\eta_{2},2d;y\right)$ terminates, and the bound states energies are obtained from the relation
\begin{equation}
d+k_{b}^{\prime }-\sqrt{k_{b}^{\prime 2}+\lambda _{K}^{2}-\eta ^{2}}=-n.
\label{ekp}
\end{equation}%
Solving (\ref{ekp}) for $k_{b}^{\prime}$, we obtain
\begin{equation}
k_{b}^{\prime }=\frac{1}{2d+2n}\left( \lambda _{K}^{2}-\eta ^{2}-\left(
d+n\right)^{2}\right) >0. \label{klb}
\end{equation}%
Using Eq. (\ref{nu1}) and the value of $k_{b}$ given in Eq. (\ref{seq2}), we find the eigenvalues
\begin{equation}
\mathcal{E}_{nl}=-\frac{\hbar ^{2}\alpha^{2}\delta ^{2}}{2M}\frac{\left(\lambda
_{K}^{2}-\eta ^{2}-\left( d+n\right) ^{2}\right)^{2}}{\left( d+n\right)^{2}}.  \label{eel}
\end{equation}
\begin{figure}[!t]
\centering
\includegraphics[width=\columnwidth]{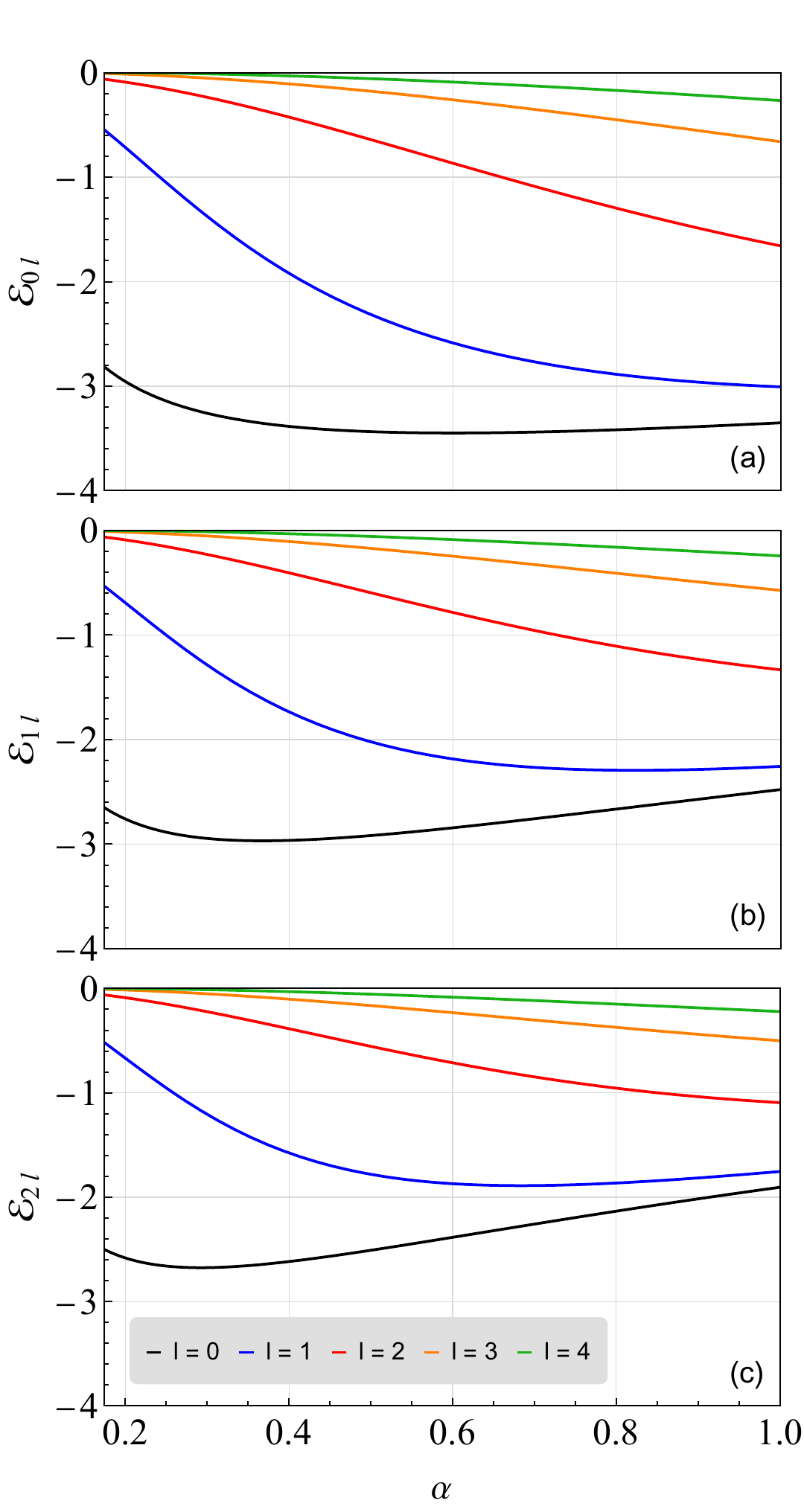}
\caption{Energy levels as a function of the monopole parameter $\alpha$ for different values of $l$. In (a), for $n=0$, (b) for $n=1$, and (c) $n=2$.
In both cases, we using $A=2$, $D=4$,  $q=1$, $\hbar=1$, $M=1$ and $\delta=0.001$.}
\label{Energy_Case_2}	
\end{figure}
\begin{figure}[!t]
\centering
\includegraphics[width=\columnwidth]{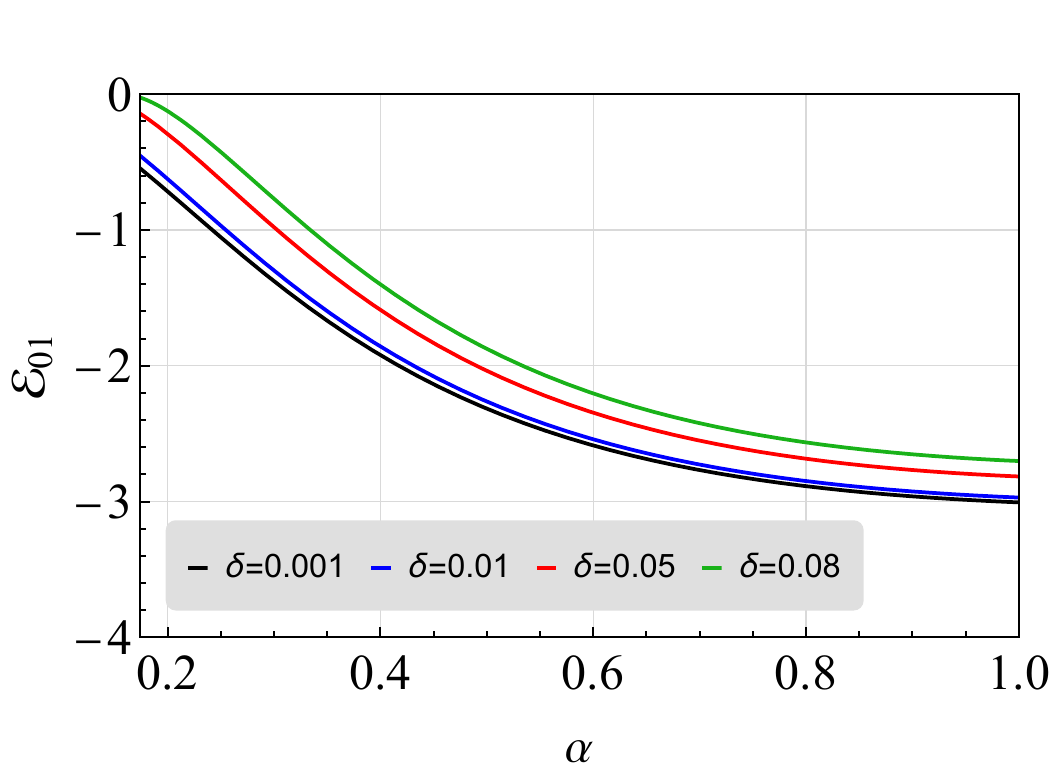}
\caption{Sketch of $\mathcal{E}_{01}$ as a function of the monopole parameter $\alpha$ for different values of $\delta$. We use $A=2$, $D=4$,  $q=1$, $\hbar=1$ and $M=1$.}
\label{Energy_Case_2_delta}	
\end{figure}
To ensure the validity of (\ref{eel}), we must remember that $k_{b}^{\prime}>0$ (see Eq. (\ref{seq2})), which requires the numerator of (\ref{klb}) to be positive. In other words, we must impose that
\begin{equation}
\lambda _{K}^{2}-\eta ^{2}-\left( d+n\right) ^{2}>0, \label{inq}
\end{equation}%
which establishes the following condition for $n$:
\begin{equation}
n<\sqrt{\lambda _{K}^{2}-\eta ^{2}}-d.
\end{equation}  

In Figure \ref{Energy_Case_2}, it is also evident that the energy exhibits significant changes in response to variations in the monopole parameter $\alpha$. The lowest energy level is $\mathcal{E}_{00}$. However, each state demonstrates its unique behavior. For instance, states such as $\mathcal{E}_{20}$ and $\mathcal{E}_{10}$ exhibit a parabolic-like trend as $\alpha$ increases. Conversely, states like $\mathcal{E}_{04}$ and $\mathcal{E}_{14}$ display a nearly linear, negative decrease (for a comprehensive view of all the curve behaviors, see Figure \ref{Energy_Case_2}). Here, also the energy levels become more compact when the quantum number $n$ increases. Nevertheless, the stabilization effect observed in the previous section for values of $\alpha$ close to $1$ is not noticeable here. Hence, the energy curves, in this case, become less predictable. Another crucial distinction between the cases with Kratzer (plus self-interaction potential) and the screened modified Kratzer potential (plus screened modified self-interaction potential) is the expanded range of the topological factor $\alpha$ due to the introduction of the screening parameter, as evidenced by the analysis of the energy spectrum behavior.
Figure \ref{Energy_Case_2_delta} demonstrates how an individual energy state can vary with the screening parameter $\delta$. We have reproduced the energy state $\mathcal{E}_{01}$ for four different values of $\delta$, revealing a shift effect where the energy becomes progressively less negative as the value of $\delta$ increases.

\section{Conclusions}
\label{sec:conclusions}

In this manuscript, we have solved the problem of a charged particle interacting with different Kratzer-type potentials in global monopole spacetime. A distinguishing factor in our work is that we take into account the self-interaction potential that the particle acquires due to the spacetime topology.

Focusing on the implications of spacetime topology and the Kratzer potential, we examine the distinct physics of the system. As a result, the energy spectrum exhibits specific behaviors and responds differently to the presence of the topological defect. Here, the bound state solutions of the radial wave equation are found in terms of the confluent hypergeometric function.

Subsequently, we  have considered a screened modified Kratzer potential, which yield significant differences in the effective potential.  In this context, we also have considered the screening effect on the self-interaction potential, which provides a full comprehensive description of the system.  Particularly, in this second case, the behavior of the energy spectrum becomes more complex and the allowed range of the topological factor offering a broader range for $\alpha$. We have used the Frobenius method and the canonical hypergeometric equation was found as the radial wave equation, which was later solved to obtain bound state solutions.

Overall, this study contributes to our understanding of the significant effects on the motion of charged particles when subjected to topological defects and interacting through molecular potentials (with or without screening modification) of the same class, particularly within the framework of a global monopole.

\section*{Acknowledgments}

This work was partially supported
by the Brazilian agencies CAPES, CNPq, and FAPEMA. E. O. Silva acknowledges CNPq
Grant 306308/2022-3, FAPEMA UNIVERSAL-06395/22. F. S. Azevedo acknowledges CNPq Grant No. 150289/2022-7. C. Filgueiras acknowledges FAPEMIG Grant No. APQ 02226/22.
This study was financed in part by the Coordena\c{c}\~{a}o de
Aperfei\c{c}oamento de Pessoal de N\'{\i}vel Superior - Brasil (CAPES) -
Finance Code 001.

\end{document}